\documentstyle[aps,preprint,prl,psfig,amssymb]{revtex}

\draft
\tighten
\begin{document}
\bibliographystyle{prsty}

\title{Structure and Dynamics of amorphous Silica Surfaces}
\author{Alexandra Roder, Walter Kob\footnote{Author to whom correspondence
should be addressed to. Permanent address: Laboratoire des Verres,
Universit\'e Montpellier II, 34095 Montpellier, France, e-mail: 
kob@ldv.univ-montp2.fr}, and Kurt Binder}

\address{Institute of Physics, Johannes Gutenberg-University,
Staudinger Weg 7, D--55099 Mainz, Germany}
\date{November 19, 2000}
\maketitle

\begin{abstract}

We use molecular dynamics computer simulations to study the equilibrium
properties of the surface of amorphous silica. Two types of geometries
are investigated: i) clusters with different diameters (13.5\AA, 19\AA,
and 26.5\AA) and ii) a thin film with thickness 29\AA. We find that the
shape of the clusters is independent of temperature and that it becomes
more spherical with increasing size. The surface energy is in qualitative
agreement with the experimental value for the surface tension.  The
density distribution function shows a small peak just below the surface,
the origin of which is traced back to a local chemical ordering at the
surface. Close to the surface the partial radial distribution functions
as well as the distributions of the bond-bond angles show features
which are not observed in the interior of the systems. By calculating
the distribution of the length of the Si-O rings we can show that these
additional features are related to the presence of two-membered rings at
the surface. The surface density of these structures is around 0.6/nm$^2$
in good agreement with experimental estimates. From the behavior of
the mean-squared displacement at low temperatures we conclude that at
the surface the cage of the particles is larger than the one in the
bulk. Close to the surface the diffusion constant is somewhat larger
than the one in the bulk and with decreasing temperature the relative
difference grows. The total vibrational density of states at the surface
is similar to the one in the bulk. However, if only the one for the
silicon atoms is considered, significant differences are found.

\end{abstract}

\vspace{-0.3cm}

\section{Introduction}

In recent years the structure and dynamics of glass-forming
systems has been the focus of many experimental and theoretical
investigations~\cite{vigo98,trieste99}. These efforts resulted in a
current understanding of such systems which is significantly larger
than it was just one decade ago, although many properties of glassy
systems still remain puzzling. Although so far most studies focused
on the bulk properties of these materials, it has become clear that
also the surfaces of glassy systems are a very interesting object
of research~\cite{pantano89,grenoble_confit_2000}. Most studies in
this direction have so far concentrated on thin polymer films or
the system studied here: silica. Whereas the former type of system
is, e.g., frequently applied in coatings, it has been found that
amorphous silica surfaces are very reactive and thus can be used in
catalysis and chromatography~\cite{legrand98,iler79,helms93}. In
addition, the properties of such surfaces are also very
important in modern materials such as reinforced polymer matrix
composites, in metal oxide semiconductor devices, or ultra-light
weight nanoporous materials~\cite{conley93,stahlbush93}. The
wide range of these applications has led to numerous
studies of the surface using a wide range of methods such as
infrared and Raman spectroscopy, atomic force microscopy, as well as
NMR~\cite{morrow76,michalske84,bunker89a,dubois93b,dubois93,grabbe95,radlein95,raberg98,gupta00,surface_studies}.
These investigations have led to the hypothesis that these surfaces
have structural elements that are not present in the amorphous bulk,
namely two-membered rings, i.e.  two oxygen atoms that are bonded to
the same two silicon atoms. Evidence has been given, that it is these
short membered rings that form the reaction sites of the surface (in
particular with water), thus making them particularly interesting.

Due to the nature of the experiments done to investigate the
surfaces, it has so far not been possible to determine the details
on the structure of these rings. Therefore efforts were made to use
various types of {\it ab initio} calculations in order to determine their
geometry~\cite{michalske84,bunker89a,kudo84,okeeffe85,hammann97,ceresoli00,lopez00,ab_initio_calc}.
However, these calculations were usually done for very small clusters
or crystalline materials. Since the influence of the surrounding on
the geometry of the rings is not understood well, the validity of
the so obtained results is unfortunately not very clear. Furthermore
these types of calculations have the problem that only very short
time scales are accessible (a few ps) and thus the systems cannot be
equilibrated at low temperatures. A different approach to investigate
the properties of such surfaces is to use classical simulations (Monte
Carlo or molecular dynamics (MD)) since they allow to access times
scales that are on the order $10^3-10^4$ times longer than the one
from the quantum mechanical calculations. In order to obtain reliable
results from such a simulation it is, however, important to have a
potential at hand which is able to describe the effective interactions
between the ions reliably. That such types of simulations are indeed
possible has been demonstrated in the seminal papers by Garofalini and
coworkers~\cite{garofalini83,garofalini85,levine87,feuston88,feuston89,feuston90,litton97}.
In these simulations a potential has been used which was able to
describe well several properties of amorphous silica in the bulk. One
of the main results of these simulations was that at the surface the
frequency of short rings is indeed much larger than the one in the bulk,
in agreement with the (indirect) evidence from the experiments. However,
due to the way the surfaces in these simulations were created (removing
of the boundary conditions in one direction {\it at low temperatures} in a
three-dimensional simulation box), they were not really relaxed, but only
annealed. Therefore it was argued that the structure of these surfaces
would not correspond to the real ones where a liquid surface is cooled
and hence can relax~\cite{bakaev99b}. Further potential difficulties
for such types of simulations are discussed in Ref.~\cite{capron97}.
Despite these possible problems recently various types of simulations
have been reported in which the structure of the silica surface was
investigated as well as its reactivity with water or other small
molecules~\cite{bakaev99b,bakaev99,beckers00,surf_sim}. What is still
lacking, however, is a simulation in which this surface is studied {\it in
equilibrium}, since to our knowledge all previous simulations considered
only annealed surfaces. Therefore we will investigate in the present
paper the structure of such surfaces and also characterize the dynamics
of the atoms. The potential we use is the one proposed some time ago by
van Beest {\it et al.}~\cite{beest90} which has been shown to be very
reliable to describe the properties of amorphous silica in the bulk. Below
we will argue that this type of potential should also be able to give
reasonable results in the vicinity of a surface.  Since the structural
and dynamical properties of this system in the bulk has in the past
been very well characterized, we can make a careful comparison of these
properties with the ones of the surface. This comparison will in turn
allow us to understand better how the surface influences these properties.

The remaining of the paper is organized as follows: In the next section
we describe the model of van Beest {\it et al.} and give the details on
the simulations. In section~III we present the results and in the last
section we summarize and discuss them.

\section{Model and Details of the Simulations}
In this section we discuss the model and give the details of
the simulations. More details on the latter can be found in
Ref.~\cite{roder_diss}.

To gain insight into the structure and the dynamics of a real surface
and the transition region between the surface and the bulk by means
of computer simulations, it is necessary to have a potential which
is able to describe reliably this type of geometry. For the case of
silica van Beest, Kramer, and van Santen (BKS)~\cite{beest90} proposed
some time ago a potential which turned out to be remarkably good in the
description of {\it bulk properties} of crystalline as well as amorphous
SiO$_2$~\cite{vollmayr96a,horbach99c,horbach99,horbach_boson,bks_sim}.
Although it is presently not clear to what extend this potential is
also reliable in the vicinity of a silica surface, the approach used
by BKS to obtain their potential makes one hope that it is also quite
reasonable in such a situation, as can be understood as follows: BKS made
self-consistent Hartee-Fock calculations to determine the energy surface
of {\it one} $H_4$SiO$_4$ tetrahedron, used a certain functional form
(described below) to parametrise this surface, and then lattice dynamics
calculations to improve further their fit parameters. Thus the initial
step in the derivation of the BKS-potential involved also the use of a
free surface, although one saturated with hydrogen, and thus it is not
unreasonable to expect that the {\it salient} features of a real silica
surface are reproduced by this potential as well. Note that this way
to obtain an effective potential is very different from the often used
one which involves fitting macroscopic data (density, elastic constants,
etc.) and microscopic data (atomic distances, bond-bond angles, etc.) from
the {\it bulk}~\cite{feuston88,vashishta90}, since in the latter approach
no surface data is considered.

The functional form of the BKS-potential is given by~\cite{beest90}

\begin{equation}
\phi_{\alpha\beta}(r) =
\frac{q_{\alpha} q_{\beta} e^2}{r} +
A_{\alpha \beta} \exp\left(-B_{\alpha \beta}r\right) -
\frac{C_{\alpha \beta}}{r^6}\quad \alpha, \beta \in
[{\rm Si}, {\rm O}],
\label{eq1}
\end{equation}

where $r$ is the distance between the ions $\alpha$ and $\beta$. The
value of the parameters $q_{\alpha}$, $A_{\alpha \beta}$, $B_{\alpha
\beta}$, and $C_{\alpha \beta}$ can be found in Ref.~\cite{beest90}. In
order to get a better description of the density in the bulk,
the short range part of the potential was truncated and shifted at
5.5\AA~\cite{vollmayr96a}. 

With this model two types of simulations were done: A first one in
which a free cluster of silica was studied and a second one in which
a thin film was considered, i.e. we used periodic boundary conditions
in the $x-$ and $y-$direction and an infinitely large space in the
$z-$direction~\cite{brangian99}. Three different cluster sizes were
considered: $N=436$, 1536, and 4608.  To generate these clusters we
proceeded as follows: We first placed {\it randomly} the $N$ ions of
SiO$_2$, in stoichiometric composition, in a sphere of radius $R_w$
at the origin of the coordinate system.  The value of $R_w$ was
chosen to be substantially larger than a sphere with $N$ ions at the
density of amorphous silica (approx. 2.2g/cm$^3$~\cite{mazurin83}),
i.e. $R_w=10$, 20, and 30\AA~for the three system sizes. Due to the
randomness, this configuration is very unstable and the ions will
accelerate quickly and would escape the spherical region in short
time. In order to prevent this we added an external potential at $R_w$
of the form $V_w(r) = (R_w-r_i)^{12}$, where $r_i$ is the distance of
particle $i$ from the origin. (The details of this external potential are
completely irrelevant, since it is used only at the early stage of the
equilibration process.) The equations of motion were then integrated with
the velocity form of the Verlet algorithm, using a time step of 1.6fs.
To equilibrate the system we coupled the ions to a stochastic heat
bath with a temperature $T$ of 4700K. This was done for 5000 time steps
which corresponds to several typical relaxation times of the system at
this temperature.  After this time, the system was equilibrated so far
that it had assumed the shape of a drop and the ions did not interact
anymore with the external potential.  Hence the latter one was switched
off and we propagated the particles for additional 10,000 steps in the
potential given by Eq.~(\ref{eq1}). Note that during this equilibration
period care was taken that the total momentum as well as the angular
momentum were always zero, in order not to put any centrifugal force on
the outer particles of the cluster and hence to change the equilibrium
structure. The temperatures investigated were 4700K, 4300K, 4000K, and
3400K.  In order to improve the statistics of the results we averaged
the smallest, medium, and large system over 10, 5 and 3 completely
independent runs. In addition we simulated the smallest systems also at
2750K (4 independent samples).

Note that with this geometry we do not have periodic boundary
conditions. Hence it is not possible to calculate the long range part
of the Coulomb forces by means of the Ewald sum, a technique in which
the computational load increases roughly like $N^{1.5}$. Instead it
was necessary to make a double loop over all pairs of ions, i.e. the
computational load increases like $N^2$. Due to this heavy load these kind
of investigations are hence only possible by using parallel computers. The
total amount of computer time spent for this studies are around 19 years
of single processor time on a Cray T3E.

For the simulation of the film we used periodic boundary conditions
in the $x$ and $y$ direction. Since the system is not periodic in the
$z$-direction, the usual Ewald summation technique to calculate the long
range Coulomb forces cannot be used in this case. Some time ago Parry and
Heyes proposed a method how such long range forces can be handled also
in a quasi-two dimensional geometry~\cite{parry,heyes} and we followed
their approach~\cite{roder_diss}. In analogy to the three dimensional
case the forces are decomposed into a sum in real space and one in
reciprocal space. Hence also here a parameter $\alpha$ and a maximum
cutoff wave-vector $k_{\max}$ occur, which we chose to be 0.265 and 6,
respectively. The number of atoms  were 1152 and in the $x$ and $y$
direction the box had a size of 28.5\AA. As a starting configuration we
used a slab of $\beta-$cristobalite and melted it at 5200K for 30,000
time steps (49ps). This time was sufficiently long to allow the ions to
move on the order of 10\AA, which is large enough to melt the system
completely.  The result was an amorphous film with a thickness around
29\AA. This system was subsequently cooled and equilibrated at 4300K
and 3200K before the production runs were started.

\section{Results}
In this section we present the results from the simulation of the
droplets. After having characterized their general shape, we focus
on comparing the structure of the network at the surface with the one in
the interior. Finally we will also discuss to what extend the dynamics
is affected by the presence of the surface.

\begin{figure}[h]
\psfig{file=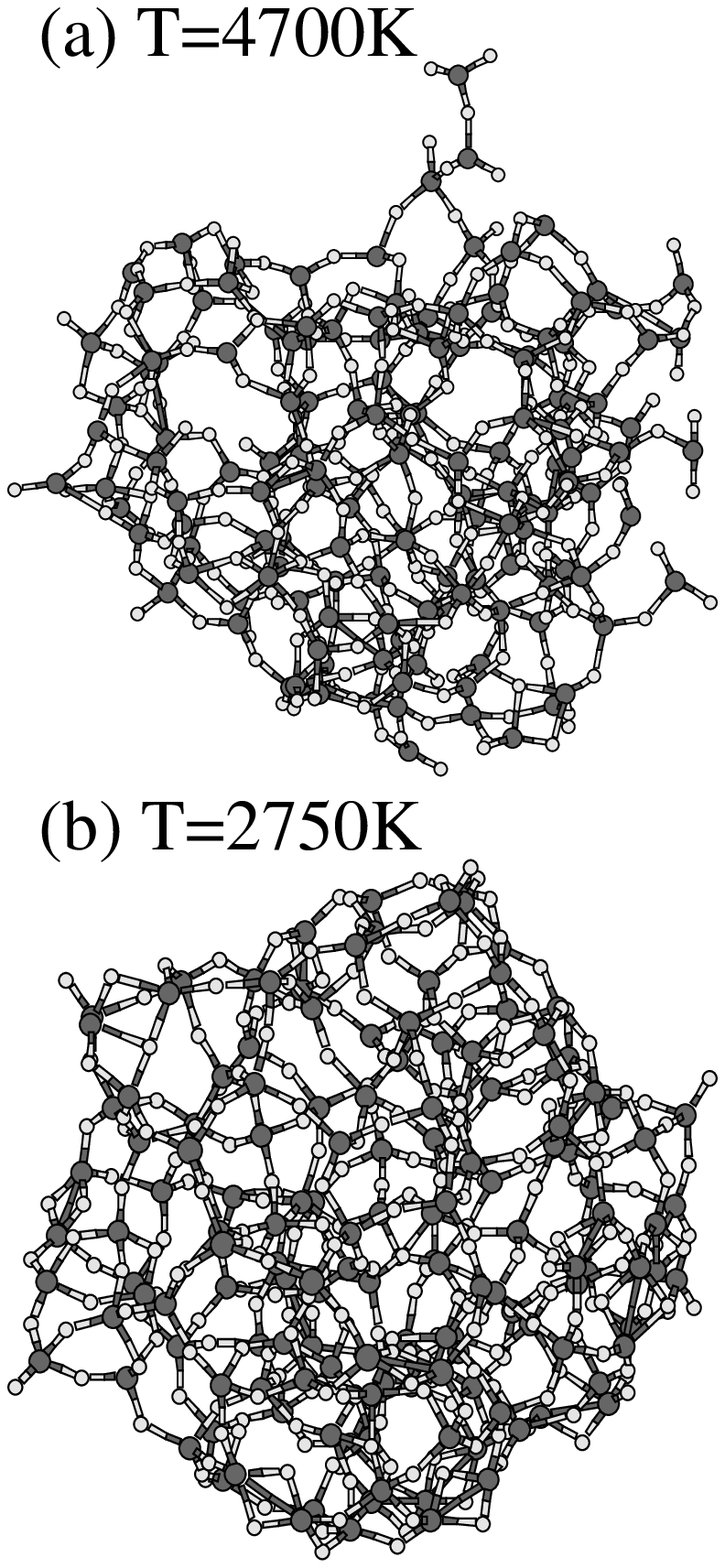,width=4.5cm,height=9.5cm}
\vspace*{3mm}
\caption{Snapshot of the system with 432 ions at two different
temperatures. The dark and light gray spheres correspond to the silicon
and oxygen atoms, respectively.}
\label{fig1}
\end{figure}

Since the size of the droplets are finite it can be expected that their
shape is not quite spherical, since fluctuations will give rise to
deviations from a sphere on the microscopic level as well as on the
length scale of the whole droplet. The size of these fluctuations will
in general depend on temperature as well as the size of the system. In
order to give an idea on the shape of the droplets we show in
Fig.~\ref{fig1} two typical snapshots of the smallest system at two
different temperatures. We see that the configuration for the higher
temperature, Fig.~\ref{fig1}a, has a surface which is very irregular in
that many arms point into the vacuum. These arms are formed of silicon
atoms that are surrounded by a few oxygen atoms. During the simulations
we found that at these high temperatures occasionally the end of such
an arm breaks off and becomes detached from the rest of the cluster.
These fragments are usually SiO$_2$ ``molecules'' and in most cases
they fall back onto the cluster at a different place they started from.
However, it is clear that occasionally such a fragment will have a
velocity high enough to leave the cluster for good and thus will not
fall back onto the cluster. Hence, on long time scales the cluster is
only metastable, since it will ultimately evaporate. (The same will be
true also in the thin film geometry.) However, on the time scale of the
simulation we have not observed these processes for $T\leq 4700$K, but
for higher temperatures occasionally a fragment is indeed spawned off
to infinity~\cite{footnote}.

If the temperature is lowered, most of the mentioned arms disappear
and the surface becomes much smoother, Fig.~\ref{fig1}b. At these low
temperatures only occasionally dangling O-bonds are sticking out from
the cluster. However, despite this local smoothness the overall shape of
the cluster is far from being spherical, since its surface is still rough
at larger length scales. (We emphasize that this roughness is {\it not}
static, since the cluster reconstructs itself on the time scale of a
few ns.) Using viscoelastic theory, J\"ackle and Kawasaki addressed the
problem of the roughness of a supercooled liquid~\cite{jackle95}. They
predict that this roughness should be given by $k_B\sqrt{T/4\pi\sigma}$,
where $\sigma$ is the surface tension. If we use $T=3000$K and a surface
tension of 0.33N/m~\cite{mazurin83}, we obtain a predicted roughness
around 1\AA, a value which in view of Fig.~\ref{fig1}b is very reasonable,
and is also in good agreement with experimental data~\cite{gupta00}.

\begin{figure}[h]
\psfig{file=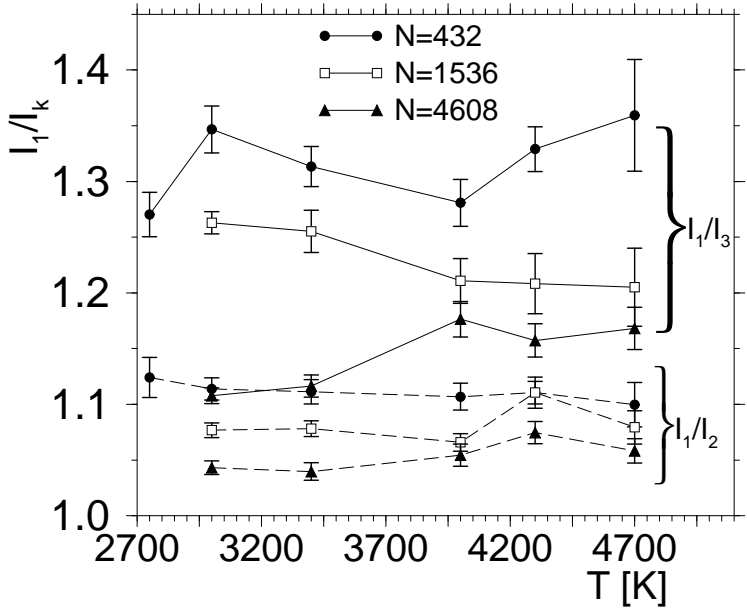,width=10cm,height=8cm}
\vspace*{5mm}
\caption{Temperature dependence of the ratio of the principal moments
of inertia, $I_1 \geq I_2 \geq I_3$, for the three system sizes.}
\label{fig2}
\end{figure}

In order to characterize the shape of the cluster, we have calculated
its three principal moments of inertia, and label them according to
their magnitude $I_1\geq I_2 \geq I_3$. In Fig.~\ref{fig2} we show
the temperature dependence of the ratios $I_1/I_3$ and $I_2/I_3$,
for the three system sizes. From this figure we recognize that the
typical shape of the cluster is rather an ellipsoid than a sphere since
the ratio of the long axis to the short one is around $\sqrt{1.32}
\approx 1.15$ for the smallest cluster and $\sqrt{1.15} \approx 1.07$
for the largest one. This decrease of the ratio with increasing $N$
can be understood as follows: If we assume that $\Delta$, the amplitude
of the capillary waves on the surface, is only a weak function of $N$,
such as e.g. a logarithmic dependence, we expect that the length of the
axis is given by $\Delta+a_1 N^{1/3}$, where $a_1$ is a slowly varying
function (with respect to $N^{1/3}$) of $N$. Hence the ratio $I_1/I_3$
will scale like $(\Delta +a_1 N^{1/3})^2/(\Delta +a_2 N^{1/3})^2$, where
$a_2$ is also a constant. If one expands this quotient in powers of
$N^{-1/3}$ one thus expects that $I_1/I_3 \approx 1+ const.  N^{-1/3}$,
and this is indeed what we find within the accuracy of our data. Note
that this reasoning holds also for the ratio $I_1/I_2$, and indeed we
find the same scaling in our data. Finally we mention that since these
ratios are independent of temperature within the accuracy of our data,
we can conclude that the amplitude $\Delta$ is essentially independent of
$T$ also.  This implies that the surface tension is only a weak function
of $T$, at least in the temperature range considered, and below we will
give some indirect evidence that this indeed the case.

Since in this paper we are interested to characterize the properties of
the surface, we need a criterion to decide which atoms belong to the
surface and which ones belong to the interior. Since the shape of the
droplets is not spherical, we approximated it by an ellipsoid which was
defined as follows. For any given droplet we constructed an ellipsoid
whose ratio of the principle moments of inertia were the same as the one
found for this droplet (assuming that the ellipsoid had a uniform mass
distribution). The length of the largest axis was chosen such that the
droplet just touched the exterior of the droplet. All atoms that were
within 5\AA~of this hull were defined to belong to the surface. All atoms
that had a distance between 5\AA~ and 8\AA~ from the hull were defined
to belong to a transition zone, and all atoms with a distance larger
than 8\AA~ were defined to belong to the interior of the droplet. In
the following we will call the atoms belonging to the interior, to the
transition zone, and to the surface also to belong to shell~1, shell~2,
and shell~3. Although this choice of the thickness of the surface and the
transition zone are somewhat arbitrary, the values are very reasonably,
as we will see below.

Finally we mention that we found that the average length of the largest
axis is 13.5\AA, 19\AA~ and 26.5\AA~ for the smallest, medium, and largest
droplet. This length depends only very weakly on temperature, since the
thermal expansion coefficient of silica is quite small. Note that this
size shows to a high accuracy a $N^{1/3}$ dependence, in agreement to
the assumption used before that the overall shape of the droplet does
not depend strongly on $N$. 

\begin{figure}[h]
\psfig{file=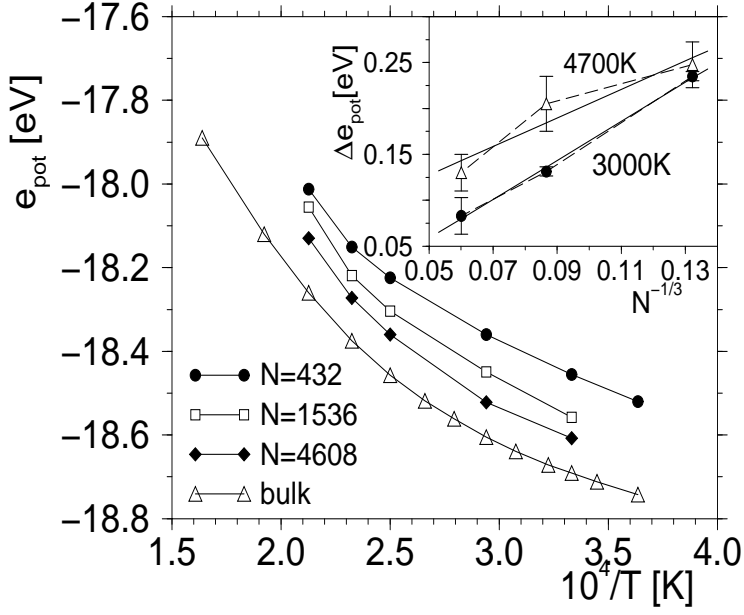,width=10cm,height=8cm}
\vspace*{5mm}
\caption{Temperature dependence of the potential energy per particle for
the three system sizes and the bulk from Ref.\protect\cite{horbach99}. 
Inset: Difference between $e_{\rm pot}(N)$ and the potential energy in 
the bulk for two temperatures.}
\label{fig3}
\end{figure}

The next quantity we consider is $e_{\rm pot}$, the total potential energy
per particle. In Fig.~\ref{fig3} we show the temperature dependence of
$e_{\rm pot}$ for the three different system sizes. Also included is the
data for the bulk, which stems from a simulation of 8016 particles in
a cubic box with periodic boundary conditions~\cite{horbach99}. We see
that the energy for the clusters is significantly higher than the one of
the bulk which is reasonable since the former include also the surface 
energy $S$. We thus expect the relation

\begin{eqnarray}
\Delta e_{\rm pot} \equiv e_{\rm pot}(N)-e_{\rm pot}^{\rm bulk} & = 
&\frac{\epsilon S}{N} 
\label{eq2}\\
& \propto & \epsilon N^{-1/3}
\quad ,
\label{eq3}
\end{eqnarray}

where $\epsilon$ is the surface energy per unit area, and in the second
equality we have assumed that the shape of the cluster is independent
of $T$, which in view of the results presented in Fig.~\ref{fig2} is
very reasonable.  Note that $\epsilon$ is {\it not} the surface tension
$\sigma$, since the latter gives the excess per unit area of the {\it
free} energy. Thus we have $\epsilon \geq \sigma$, due to the entropic
contribution to the free energy.

According to Eq.~(\ref{eq3}) the $N-$dependence of $\Delta e_{\rm pot}$
should be $N^{-1/3}$. That this is indeed the case is demonstrated in
the inset of Fig.~\ref{fig3} where we plot this quantity for two
temperatures. The value of $\epsilon$ can thus be estimated from $\Delta
e_{\rm pot}$ via Eq.~(\ref{eq2}) by assuming that the droplet is a
sphere of radius $R_0$ and hence its surface is $4\pi R_0^2$. For
$T=3000$K we find for all three system sizes $\epsilon=(0.7\pm 0.1)$N/m,
where we have taken for $R_0$ the value of the largest axis of the
above mentioned ellipsoid.

\begin{figure}[h]
\psfig{file=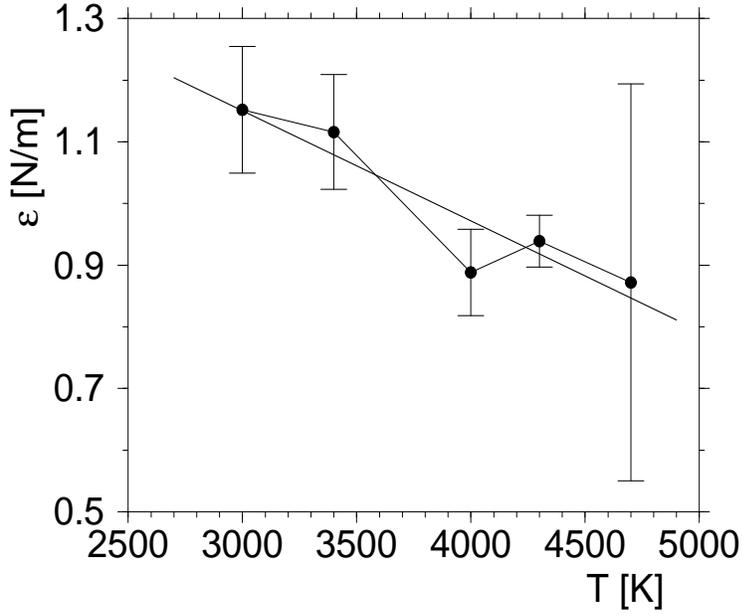,width=10cm,height=8cm}
\vspace*{5mm}
\caption{Temperature dependence of $\epsilon$, the surface energy per unit
area. The solid line is a linear fit to the data.}
\label{fig4}
\end{figure}

From Eq.~(\ref{eq2}) it is clear that the $T-$dependence of $\Delta
e_{\rm pot}$ can have two origins: The $T-$dependence of $\epsilon$
and the one of $S$. Above we have argued, using the results of
Fig.~\ref{fig2}, that this latter $T-$dependence is only weak, a
conclusion that we will confirm below. Therefore we can calculate the
$T-$dependence of $\epsilon$ from the one of $\Delta e_{\rm pot}$, see
Fig.~\ref{fig3}. The results are shown in Fig.~\ref{fig4}. We see that
$\epsilon$ decreases with increasing temperature, in agreement with
experimental findings {\it for the surface tension}~\cite{mazurin83}.
The slope that we find is $-(17\pm 9 )\cdot 10^{-3}$N/m per 100K, which
compares well with the estimate from experiments which is around $-10
\cdot 10^{-3}$N/m per 100K for $\sigma$~\cite{mazurin83}. We also
mention that the experimental value of $\sigma$ around 2000K is
$0.33\pm 0.04$N/m~\cite{mazurin83}, i.e. it is around a
factor of three smaller than an extrapolation of $\epsilon$ to this
temperature. Since we have argued above, that $\epsilon$ must be larger
than $\sigma$, we thus find that our result for $\epsilon$ is quite
reasonable.

\begin{figure}[h]
\psfig{file=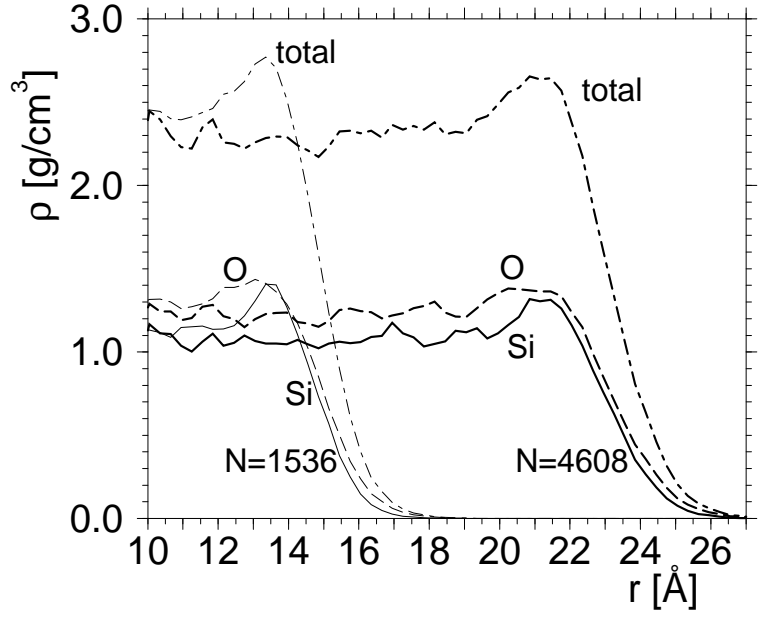,width=10cm,height=8cm}
\vspace*{5mm}
\caption{Density profile for the medium and large cluster (thin
and bold curves, respectively). Dashed-dotted lines are for the total
density. Solid and dashed lines are for the density of silicon and oxygen,
respectively.}
\label{fig5}
\end{figure}

We now investigate the structure of the clusters in more detail. One
important structural quantity is the dependence of the particle density
$\rho(r)$ on the distance $r$ from the center of the droplets. To
calculate $\rho(r)$ we determined for each configuration the principal
axes of the enclosing ellipsoid, see description above. We then
constructed a sequence of ellipsoids that were centered at the center
of gravity of the cluster, had the same geometrical shape, and whose
smallest axis increased in steps of 0.15\AA.  We then counted the
number of atoms in each shell formed by two consecutive ellipsoids and
normalize it by its volume. In Fig.~\ref{fig5} we show the so obtained
density distribution at $T=3000$K for the intermediate and largest
system (bold and thin curves, respectively). We see that at distances
that are a few \AA~ smaller than the radius of the cluster, e.g. $r\leq
18$\AA~in the case of $N=4608$, the total density is constant to within
the accuracy of the data, and is independent of $N$. This density is
around 2.3g/cm$^3$, the same value that was found for this model in the
bulk around zero pressure~\cite{vollmayr96a,horbach99}. Thus we have
a first evidence that in the interior of the cluster the structure is
similar to the one in the bulk. 

In the region close to the boundary of the cluster the total density has a
small peak. If one looks at the partial densities for silicon and oxygen
separately, solid and dashed lines, respectively, one sees that this
feature is more pronounced for the case of silicon. The reason for this
is that for the system it is energetically better to have an oxygen atom
at the surface, since in that way only one bond, if any, has to be broken
(i.e. the system forms a dangling bond), whereas if a silicon is at the
surface several bonds have to be broken. Thus one expects that the density
profile for the oxygen atoms extend to larger distances than the one for
silicon, and this is indeed what can be read off from the figure. Due
to the excess of oxygen at the surface, the tendency to achieve local
charge neutrality makes it now very likely that the layer just below the
surface has an excess in silicon, thus rationalizing the peak in the total
density. A qualitatively similar behavior has been found by Garofalini
and coworkers for a different potential~\cite{litton97} which
shows that this oxygen enrichment is not an artifact of the present
model. 

\begin{figure}[h] 
\psfig{file=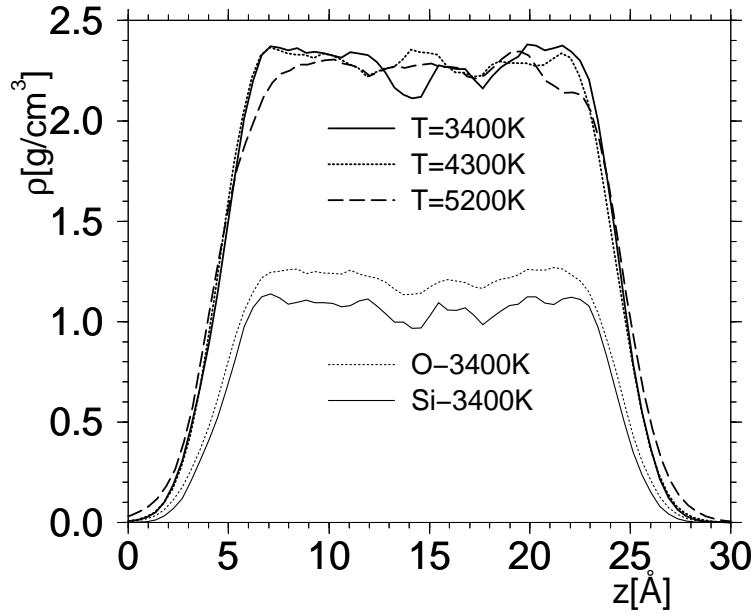,width=10cm,height=8cm}
\vspace*{5mm}
\caption{Density profile for the thin film. The upper three curves are
the total mass density at three different temperatures. The lower two
curves are the density for the oxygen and silicon atoms at $T=3400$K.}
\label{fig5bis}
\end{figure}

That this effect can also be seen in the thin film is demonstrated in
Fig.~\ref{fig5bis}, where we show the density profile for this type
of geometry. The two lower curves are the densities for the oxygen and
silicon atoms at 3400K. As in the case of the cluster, we see that the
distribution for the oxygen atoms is slightly wider than the one for the
silicon atoms. Furthermore we note also here, that the latter distribution
shows a small peak close to the two surfaces, thus signaling the presence
of an excess of silicon just below the surface. 

The curves shown in Fig.~\ref{fig5} are for $T=3000$. If the temperature
is increased, the height of the flat region at intermediate and
small values of $r$ does not change, which is in agreement with
the fact that the bulk density of silica is only a weak function of
temperature~\cite{mazurin83}, a property that is reproduced well by
the present model~\cite{vollmayr96a,horbach99}. However, the density
profile very close to the surface does depend on temperature in that with
increasing temperature the height of the excess peak diminishes and the
final decay to zero at large distances becomes less steep. This effect is
demonstrated for the film geometry in Fig.~\ref{fig5bis}, where we show
the distribution of the total density for three different temperatures
(top three curves). This behavior can easily be understood by recalling
that at high temperatures the cluster has many small thin arms that
contribute to the density also at large distances (see Fig.~\ref{fig1})
and hence make the surface less well defined. Since these results for
the surface of the clusters and the ones of the films are so similar,
and this is the case also for many other ones~\cite{roder_diss}, we will
in the following not discuss the latter ones anymore.

\begin{figure}[h]
\psfig{file=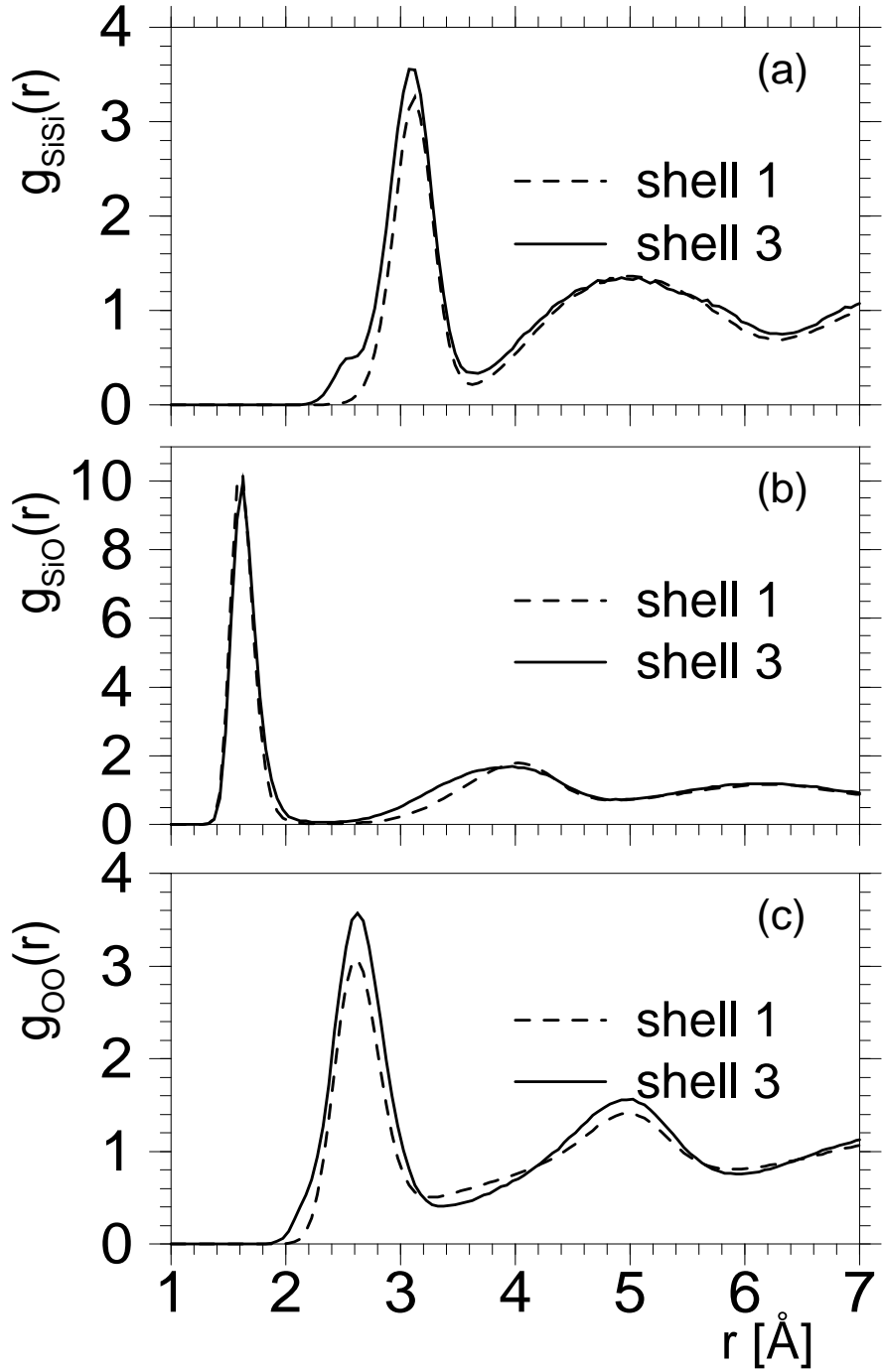,width=12cm,height=18cm}
\vspace*{5mm}
\caption{Partial radial distribution functions $g_{\alpha\beta}(r)$
for the cluster with $N=4608$ ions at $T=3000$K. Shell~1 and 3 
correspond to the interior and surface of the cluster, respectively. 
a) $g_{\rm SiSi}(r)$, b) $g_{\rm SiO}(r)$, c) $g_{\rm OO}(r)$.}
\label{fig6}
\end{figure}

A quantity that gives information on the structure on a more microscopic
level than the density profile is the (partial) radial distribution
function $g_{\alpha\beta}(r)$~\cite{hansenmcdonald86}, which gives
the probability of finding an atom of type $\alpha$ at a distance $r$
from an atom of type $\beta$, normalized by the phase space factor. In
a infinite system (or one with periodic boundary conditions), this
factor is just $4\pi r^2$. For an inhomogeneous system this factor
depends on the position of atom $\alpha$ and the resulting expression
is somewhat cumbersome, but simple to derive. Therefore we refer
the interested reader to Ref.~\cite{roder_diss} for more details on
this. In Fig.~\ref{fig6} we show the resulting $g_{\alpha\beta}(r)$ for
the largest system at $T=3000$K. In each panel two curves are shown:
The solid ones correspond to the $g_{\alpha\beta}$ if one of the atoms
is in the surface shell (shell 3).  The dashed line is for the atoms
that are in the interior of the cluster (shell 1). We mention that
within the accuracy of our data the curves of shell 1 are identical to
the ones of the bulk~\cite{horbach99}. This shows that with respect to
this observable the interior of the cluster is equivalent to the bulk.

From the figure we recognize that the $g_{\rm SiO}(r)$ in shell~1 and
shell~3 are quite similar. In particular the first nearest neighbor
peaks are almost identical, which shows that the surface does not
change the nearest neighbor distance in Si-O. This can be understood
by recalling that this type of bond is very strong and hence it does
not depend on the environment. The second nearest neighbor peak of
$g_{\rm SiO}$ is, however, a bit broader than the one in the bulk,
which shows that on the surface the system is more heterogeneous than
inside. Also the distribution $g_{\rm OO}(r)$ in shell~3 is quite
similar to the one in shell~1, although in this correlation function
also differences in the first peak are observed in that on the surface
the peak is higher and broader. The largest differences are found in
$g_{\rm SiSi}(r)$, where we see that around 2.5\AA~a small shoulder is
observed in the curve for the surface. This means that in this shell
a new length scale exists, the origin of which will below be shown to
be two-membered rings, i.e. a structural unit which is not found in
the bulk~\cite{vollmayr96a,rino93}. Finally we mention that we have
also calculated the three partial structure factors for the clusters
and found that these functions do not show any unexpected behavior or
give new insight into the structure of the droplets, beyond the one
of $g_{\alpha\beta}(r)$.

\begin{figure}[h]
\psfig{file=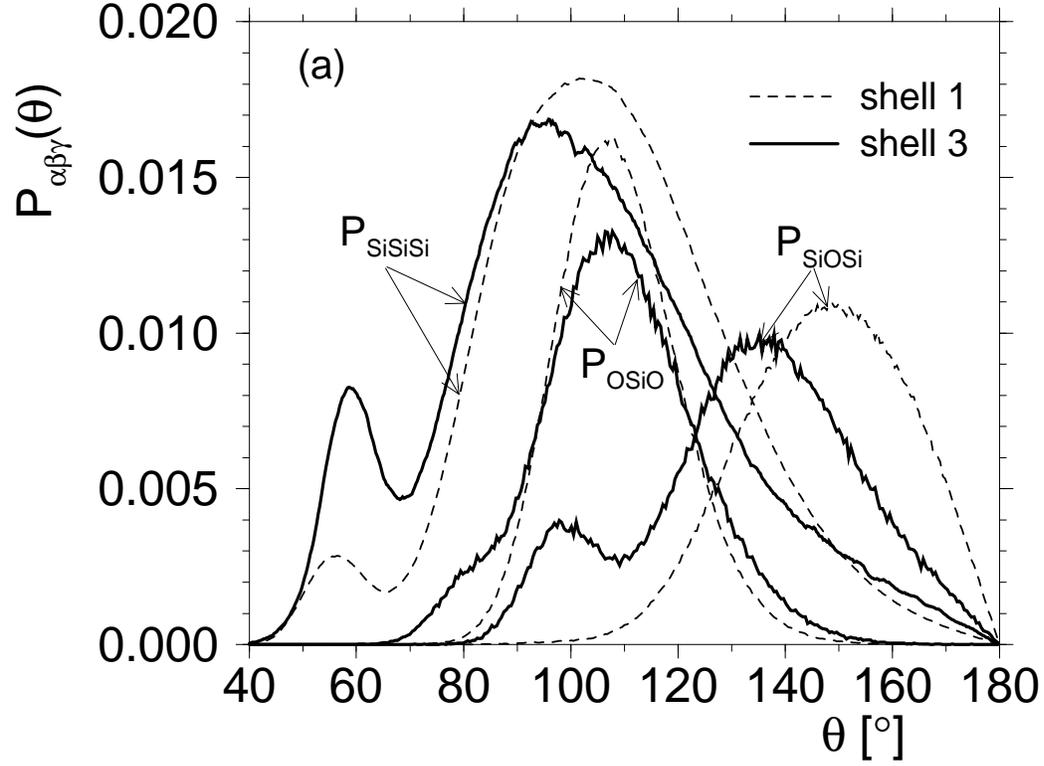,width=14cm,height=10cm}
\vspace*{5mm}
\psfig{file=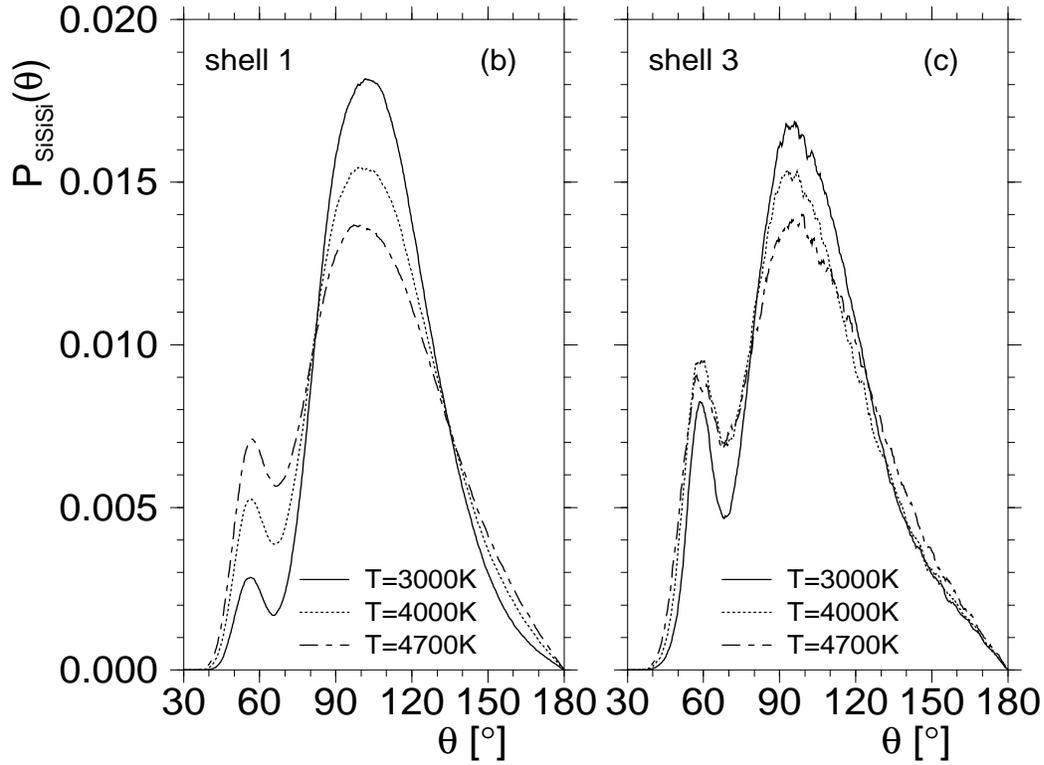,width=14cm,height=10cm}
\vspace*{5mm}
\caption{Probability that an angle between neighboring atom of type
$\alpha$, $\beta$, and $\gamma$ has a value $\theta$ for the largest
system, $N=4608$, $T=3000$K. a) shell~1 (interior) and shell~3 (surface); 
b) and c) temperature dependence for shell~1 and shell~3, respectively.}
\label{fig7}
\end{figure}

Further relevant information on the local structure can be obtained
from the distribution function of the angle between three neighboring
particles.  For this we define two particles (of type $\alpha$ and
$\beta$) to be neighbors if their distance is smaller than the first
minimum in $g_{\alpha\beta}(r)$. These minima are at 3.63\AA~, 2.32\AA,
and 3.24\AA, for the Si-Si, the Si-O, and O-O correlation function,
see Fig.~\ref{fig6}, and are independent of temperature. Some of the
angle distribution function $P_{\alpha\beta\gamma}(\theta)$ are shown in
Fig.~\ref{fig7}a. These data are for the largest system at $T=3000$K,
and we have found that for the smaller systems the results are very
similar. Again we show for each distribution two curves: One for the
innermost shell and one for the surface. (Also for this observable the
distributions for the interior of the cluster is essentially identical to
the one of the bulk.) In contrast to the quantities investigated so far,
we see in this observable a very strong difference between the structure in
the bulk and the one close to the surface. E.g. the main peaks in $P_{\rm
SiSiSi}$ and in $P_{\rm SiOSi}$ for the surface are shifted by about
10$^{\circ}$ to smaller angles and have become broader. More important,
the latter distribution shows now a second peak around 100$^\circ$, a
feature which is not present at all in the bulk. A similar new feature
is found in $P_{\rm OSiO}$, where around 80$^{\circ}$ a shoulder is
observed which is not present in the bulk. Although $P_{\rm SiSiSi}$
shows a secondary peak also in the bulk, its amplitude is much larger
in the surface layer.

In Fig.~\ref{fig7}b~and c we show the temperature dependence of
the distribution function $P_{\rm SiSiSi}$ for shell~1 and shell~3,
respectively. From panel (b) we see that the secondary peak around
60$^{\circ}$ mentioned above shows a strong temperature dependence in
that its amplitude decreases rapidly with decreasing temperature. In
Ref.~\cite{vollmayr96a} is was shown that (in the bulk) even at
temperature zero a small peak at this angle is observed, but that its
amplitude decreases if the cooling rate with which the sample is quenched
to low temperatures decreases. Thus those results are in agreement
with the present ones. For the case of the surface, the temperature
dependence of the amplitude of the secondary peak is much weaker, see
panel (c). Thus it is reasonable to assume that this structural feature
will be present even at low temperatures, in contrast to the peak in the
bulk. A qualitatively similar behavior is found for the secondary peak
in $P_{\rm SiOSi}$. Thus we conclude that the surface has structural
elements that are present in the bulk only at high temperatures, whereas
on the surface they are found at all temperatures. These results are in
qualitative agreement with the (less detailed) studies of Garofalini on
these distributions~\cite{garofalini83}.

In order to find out the nature of these structural elements we
investigated the distribution of the length of the ``rings'' in the
network. For this we define a ring as follows: Start from any silicon
atom and pick two of its nearest neighboring oxygen atoms (neighbors
are defined again by means of the minimum in the radial distribution
function). Find the {\it shortest} consecutive sequence of Si-O elements
that connect these two oxygen atoms. In this way we have constructed
a closed loop of Si-O segments and the length of the ring is defined
as the number of silicon atoms in this loop.  Note that if an oxygen
is attached only to one Si atom, i.e. a non-bridging oxygen, we define
this length to be 1. The interest in these rings stem from the fact that
they give information on the structure on intermediate distances. Also,
since the connectivity of silicon atoms can (indirectly) be studied
in NMR experiments (see, e.g., Ref.~\cite{stebbins95} and references
therein), they are not only of theoretical, but also of experimental
relevance, thus justifying the investigations on them done in the
past~\cite{vollmayr96a,rino93}.

\begin{figure}[h]
\psfig{file=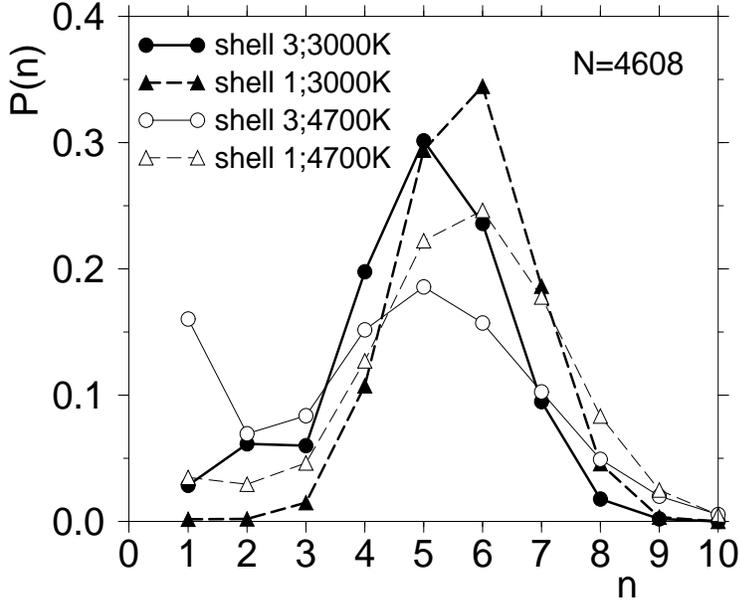,width=10cm,height=8cm}
\vspace*{5mm}
\caption{Probability that a ring has length $n$ for the case that the
ring is close to the surface (circles) and is in the interior (triangles).}
\label{fig8}
\end{figure}

In Fig.~\ref{fig8} we show the distribution function $P(n)$, i.e. the
probability that a ring has length $n$. The data is for the largest
system and we show the distribution for the interior and the surface at
two temperatures. (A ring belongs to the surface if at least one silicon
atom is in shell~3.) From the graph we see that the most probable length
of a ring is 5 or 6, a result that can be understood by recalling that
on the local length scale the structure of the melt at zero pressure
will be very similar to the one of $\beta-$cristobalite~\cite{stoffler69},
a crystalline form of SiO$_2$  that has only 6 membered rings. Note that
the whole distribution for the surface is shifted a bit to smaller values
of $n$. This effect can be understood by realizing that if an atom is
close to the surface it is more difficult for it to be part of a large
ring, due to entropic reasons.

At high temperatures, $T=4700$K, one sees that also the above mentioned
dangling bonds, $n=1$, are quite frequent, in particular at the
surface. This is in qualitative agreement with the configuration snapshot
of Fig~\ref{fig1}.  The number of the mentioned dangling bonds depends
strongly on temperature in that a change of $T$ from 4700K to 3000K gives
rise to a decrease by a factor of 4-5 in the case of the surface and in
the interior they practically vanish completely. For the two-membered
rings the situation is different. Although in the interior the change of
temperature leads again to a very strong reduction of their frequency,
their number decreases only slightly in the case of the surface. Also
rings of length three are, at the lower temperature, roughly three times
more frequent on the surface than in the bulk. Thus we conclude that close
to the surface there is a substantial fraction of very short rings. Due to
their shortness, the angles occurring in such rings are relatively small,
which explains our findings on the distribution of the bond-bond angles,
Fig.~\ref{fig7}, that in the outermost shell peaks at relatively small
angles appear. These results are in qualitative agreement with the
ones found earlier by Feuston and Garofalini~\cite{feuston89},
although the relative intensities are quite different.

\begin{figure}[h]
\psfig{file=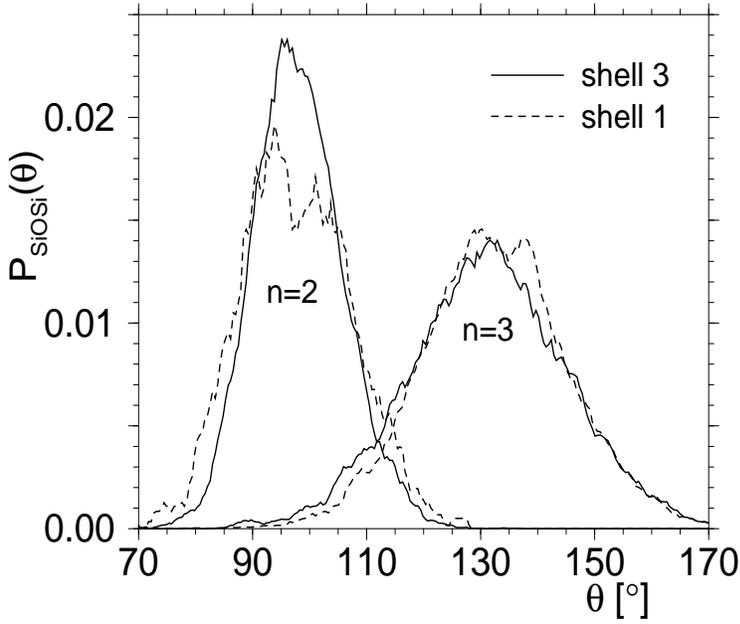,width=10cm,height=8cm}
\vspace*{5mm}
\caption{Probability that in a ring of length $n$ the angle Si-O-Si has
a value $\theta$. The solid and dashed curves are for the surface and
the interior of the cluster, respectively. $T=3000$K.}
\label{fig9}
\end{figure}

We now investigate the question whether the presence of a surface changes
only the relative frequency with which a ring of a given length appears,
or whether also the properties of the ring changes. In Fig.~\ref{fig9}
we show the distribution of the Si-O-Si angles occurring in a ring with
length $n=2$ and $n=3$. The solid and dashed lines are for the rings in
the outermost shell and the interior of the cluster, respectively. We see
that the angle for two-membered rings has a peak around 95$^{\circ}$. This
angle corresponds to the location of the secondary peak that we found
in the total distribution function in Fig.~\ref{fig7}a. Thus we can
conclude that this peak stems from two-membered rings. (Note that in
Fig.~\ref{fig7}a the curve for the interior was very close to zero around
95$^{\circ}$. The reason why we see now in Fig.~\ref{fig9} nevertheless
a peak is that {\it a few} two-membered rings are present anyway. Since
there are only a few of short rings, the curves for the bulk are quite
noisy.) From Fig.~\ref{fig9} we see that within the accuracy of our
data the distribution of the angles is the same in the interior and the
surface of the system. The same is found for the angle O-Si-O as well as
for longer rings. Therefore we conclude that the structure of the rings
is independent of their location and the difference between the surface
and the interior is only the frequency of the various rings.

In Tab.~\ref{tab1} we give the geometry of the two-membered rings as
found in the present simulation and compare it with values from another
classical simulation and quantum mechanical calculations. From this list
we see that although our simulation underestimates and overestimates
the O-Si-O and Si-O-Si angles, respectively, as obtained by the {\it
ab initio} calculations by around 7-8$^{\circ}$, the various distances
are reproduced well. Regarding the discrepancies in the angles it has to
be realized that the quantum mechanical calculations have been done for
very small clusters of SiO$_2$, thus structures that were not part of a
larger network. That such a network can have a significant effect on the
form of the two-membered rings can be inferred from the last line in the
table where we give the geometry of so-called ``silica-w'', a synthetic
polymorph of SiO$_2$ which contains only chains of two-membered rings. We
see that in this material the geometry of these rings is very different
from the ones found in the small molecules, thus showing the importance
of the surrounding.

\begin{figure}[h]
\psfig{file=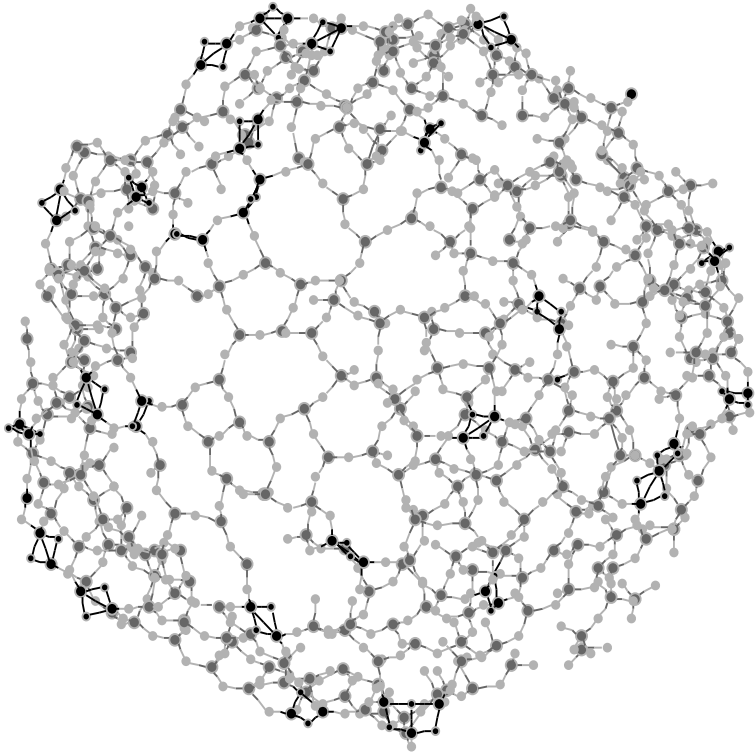,width=10cm,height=10cm}
\vspace*{5mm}
\caption{Snapshot of the $N=4608$ system at $T=3000$K. Only the top
6\AA~of the cluster are shown. Dark and light grey: silicon and oxygen
atoms. Black: Two-membered rings.}
\label{fig10}
\end{figure}

In order to get a better understanding on the geometry of the surface we
show in Fig.~\ref{fig10} a snapshot of the largest system at 3000K. All
atoms are shown that have a distance from the center that is larger than
20\AA. Since for this system size the typical radius of the cluster
is 26\AA, we thus show the top 6-7\AA, i.e. basically the shell that
we defined to be the surface. The dark and light gray atoms are the
silicon and oxygens, respectively. Marked in black are the two-membered
rings. (Note that for the sake of a better representation we show only
the upper half of the cluster.) From the figure we recognize that the
rings do not show any tendency to cluster, but rather seem to stay apart
from each other. The reason for this is likely that a region with such a
ring has a relatively high stress and thus bringing two such rings close
together is energetically rather unfavorable. We also mention that
for the largest system we find on the order of 50 two-membered rings
on the surface. With a surface area of around 8500\AA$^2$ this value
thus corresponds to an average density of 0.6 rings per nm$^2$. This
value is in good agreement with experimental data, which varies between
0.1-0.4 rings per nm$^2$~\cite{bunker89a,grabbe95,dubois93}
and other simulations which give values between 0.13-3.0 rings per
nm$^2$~\cite{feuston90,bakaev99,levine87,feuston89,ceresoli00}. Note,
however, that in Fig.~\ref{fig8} we see that the number of these rings
decreases slowly with temperature. Thus it can be expected that the
present value for the density of rings decreases if one would be able to
equilibrate the system at even lower temperatures. 

\begin{figure}[h]
\psfig{file=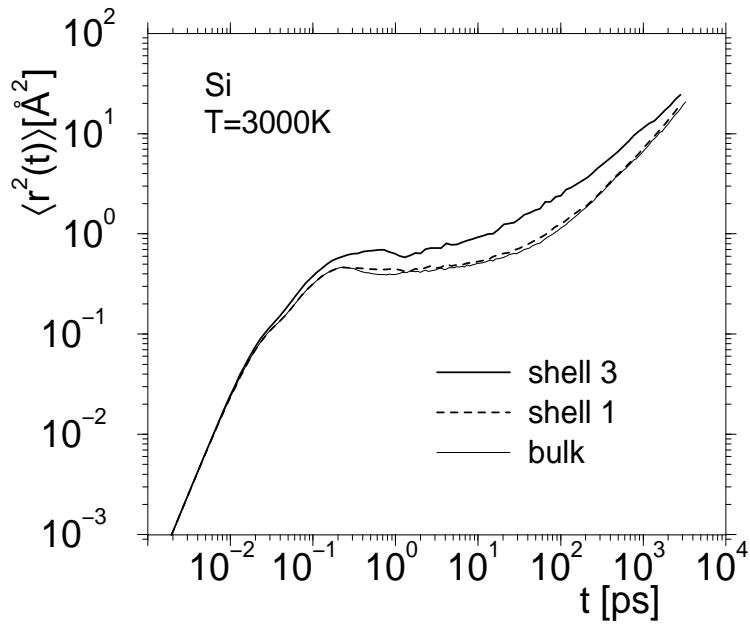,width=10cm,height=8cm}
\vspace*{5mm}
\caption{Time dependence of the mean-squared displacement of a tagged
silicon particle at 3000K. The bold solid and dashed curves correspond
to the surface and the interior of the cluster, respectively. The thin
curve correspond to the bulk~\protect\cite{horbach99}.}
\label{fig11}
\end{figure}

Having now discussed the static properties of the clusters we turn in
the final part of this section to the dynamics of this system. One of
the simplest dynamical quantities for a liquid is the mean-squared
displacement (MSD) of a tagged particle $i$, $\langle r^2(t)\rangle
=\langle |{\bf r}_i(t)-{\bf r}_i(0)|^2\rangle$, where ${\bf r}_i(t)$
is the location on the particle at time $t$. In Fig.~\ref{fig11} we
show the time dependence of $\langle r^2(t)\rangle$ for the silicon
atoms at 3000K. (The curves for the oxygen atoms look qualitatively
the same.) Three curves are shown: the ones in which at time zero
the atoms were in shell~1 or shell~3, and the curve for the bulk from
Ref.~\cite{horbach99}. Before we discuss the three curves let us briefly
comment on the various time regimes of the dynamics. At very short
times, $t\leq 0.02$ps, the particles move just ballistically, $\langle
r^2(t)\rangle \propto t^2$, since their motion is not affected by the
presence of all the other particles. At times around 0.1ps the particles
collide with their neighbors and they rattle around in the cage formed
by these neighbors. Hence the MSD is almost constant for a around two
decades in time. Only at much longer times the particles are able to
escape this cage and to become diffusive, i.e. $\langle r^2(t)\rangle
\propto t$. This sequence is the general behavior found for the MSD in
a liquid with glassy dynamics~\cite{kob99}.

From the figure we see that at short times, i.e. in the ballistic
regime, the dynamics of the atoms is independent of the shell, since
in this time window the environment does not yet affect the motion of
the particles.  The first differences between the first and third shell
are noticeable for $t \geq 0.04$ps, in that the MSD of the particles at
the surface is larger than the one of the particles in the interior. At
these times the motion of the particles has still a strong vibrational
character~\cite{horbach99} and thus we conclude that the cages seen by
the particles are larger close to the surface than in the interior of
the cluster. This result is very reasonable since we have seen above
that the atoms close to the surface have more defects and hence it
can be expected that the size of their cage is larger. Apart from the
larger size of the cages, i.e. a higher value of the plateau in the MSD,
the dynamics of the particles in the surface shell seems to be quite
similar to the one in the bulk, at least as can be inferred from this
observable (see, however, the vibrational spectrum discussed below).
Furthermore one can conclude from the graph that the dynamics of the
particles inside the cluster is very similar to the one in the bulk,
since only in the plateau regime small differences can be observed.

We also mention that we have also calculated the components of the MSD
that are parallel and perpendicular to the surface. We have found that
at low temperatures the radial component for silicon (multiplied by
1.5 to correct for the trivial geometric factor) is slightly larger
than its parallel component.  However, this effect was only small,
$O(20\%)$, and observable only in the time regime in which the MSD is
almost constant. This shows that the dynamics of the particles is almost
isotropic. The small difference can be understood by recalling that on
the surface the particles are not caged very efficiently, since there
are no caging particles towards the vacuum.

Finally we remark that the largest values of $\langle r^2(t)\rangle$
is around 22\AA$^2$, which means that on average a particle has moved a
distance around 5\AA. Thus if at time zero an atom was in a certain shell
it will usually not be able to leave this shell within the time span
of the simulation. However, for very long times the system is of course
ergodic and thus the MSD becomes independent of the shell in which the
tagged particle started in. In fact, from the figure one already sees
that for times larger than a few hundred ps the curve for shell~3 starts
to approach the one for shell~1. 

\begin{figure}[h]
\psfig{file=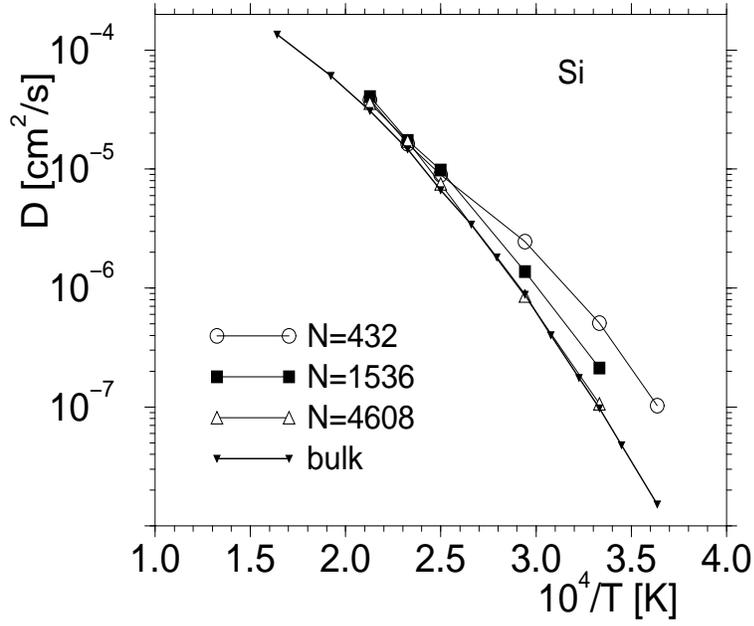,width=10cm,height=8cm}
\vspace*{5mm}
\caption{Temperature dependence of the diffusion constant for the
silicon atoms in the interior of the clusters.  The different curves
correspond to the different system sizes. Also included is the bulk data
from Ref.~\protect\cite{horbach99}.}
\label{fig12}
\end{figure}

From the MSD at long times and the Einstein relation one can immediately
calculate the diffusion constant $D$: $D=\lim_{t\to \infty} \langle
r^2(t)\rangle /6t$. (Here we have taken the average only over the particles
of shell~1.)  For the case of silicon, the resulting values of $D$ are
shown in Fig.~\ref{fig12} in an Arrhenius plot. The data for oxygen
look qualitatively similar~\cite{roder_diss}. The different curves
correspond to the various system sizes and in addition we have also
included the curve from the bulk~\cite{horbach99}. From the plot one
recognizes immediately that the relaxation dynamics of the particles
in the finite clusters is faster than the one in the bulk. However,
for the largest system this difference is only very small. We also
mention that for a strong glass former like silica~\cite{angell85},
one might expect that the temperature dependence of $D$ is an Arrhenius
law. Instead we find that {\it at intermediate and high} temperatures,
quite strong deviation from this law are present. In Ref.~\cite{horbach99}
it was argued that at these high temperatures the dynamics of the system
is qualitatively different from the one at lower temperatures since the
effects described by the so-called mode-coupling theory of the glass
transition become important~\cite{mct}. Only at low temperatures, which
for silica are presently just barely accessible in a computer simulation,
does the dynamics cross over to an activated one and thus the temperature
dependence of the transport coefficients become an Arrhenius law. We
stress, however, that the present model of silica seems indeed to be able
to reproduce reliably the dynamics at low temperatures, since, e.g.,
quantities like the activation energy of $D$ for Si and O, as well as
the one of the viscosity are reproduced well~\cite{horbach99}.

\begin{figure}[h]
\psfig{file=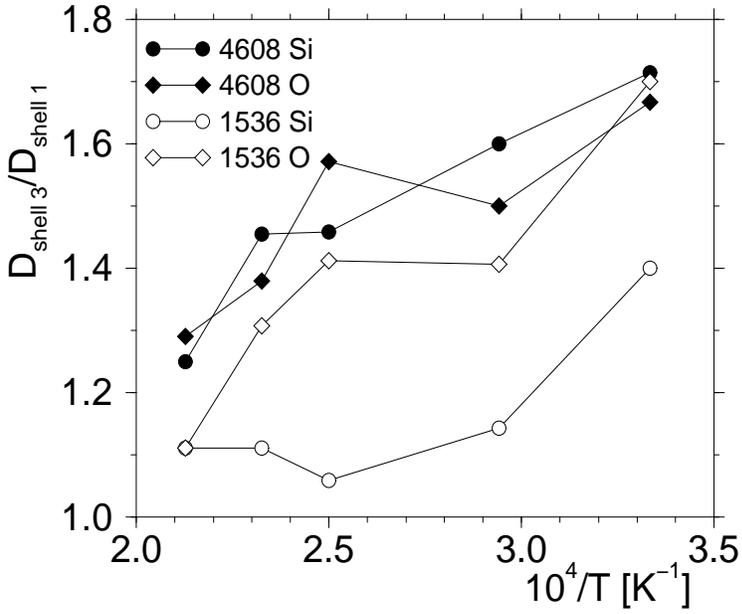,width=10cm,height=8cm}
\vspace*{5mm}
\caption{Temperature dependence of the ratio between the diffusion
constants in shell~3 and shell~1. The filled and open symbols are for
the largest and intermediate system size, respectively.}
\label{fig13}
\end{figure}

In order to show to what extend the relaxation dynamics depends on
the shell, we plot in Fig.~\ref{fig13} the ratio of the diffusion
constant For the particles on the surface and in the interior. From
this figure we recognize that at all temperatures investigated this
ratio is significantly larger than one and increases with decreasing
temperature. For the larger system, filled symbols, this ratio is
independent of the type of particle. (Although for the intermediate
system the ratio for Si and O are not the same, this is probably due to
the relatively large error for the Si data for this value of $N$.) Thus
we conclude that with decreasing temperature the motion of the particles
on the surface decouples from the ones of the particles in the interior.

\begin{figure}[h]
\psfig{file=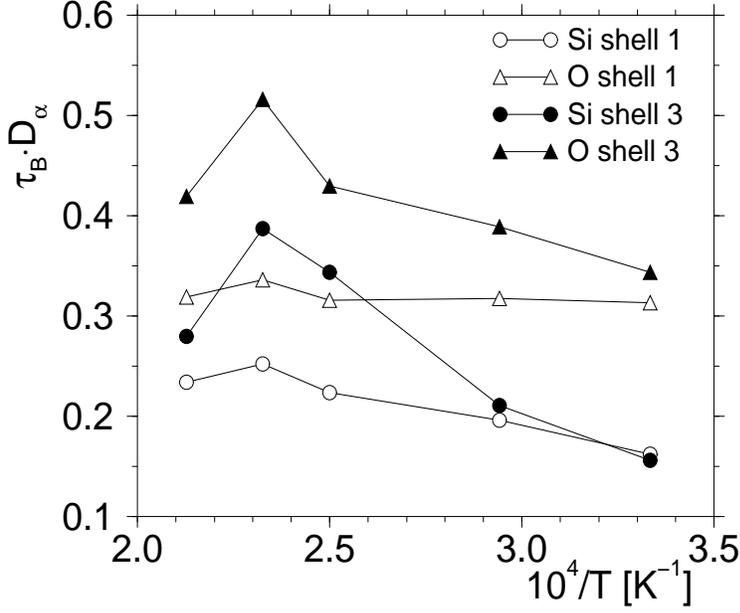,width=10cm,height=8cm}
\vspace*{5mm}
\caption{Temperature dependence of the product $\tau_B(T) D_{\alpha}(T)$
for the interior and the surface of the cluster (open and filled symbols,
respectively).}
\label{fig14}
\end{figure}

In order to understand better the reason for this decoupling it is
useful to investigate the diffusion process a bit closer. For this we
define the function $P_B(t)$, the probability that a bond between a
silicon and a neighboring oxygen atom that is present at time zero is
also present at time $t$. In the past it has been found that in the
bulk the shape of $P_B(t)$ is independent of temperature and that
therefore it is possible to define a typical life-time of a bond,
$\tau_B$, by $P_B(\tau)=e^{-1}$~\cite{horbach99}. In Fig.~\ref{fig14}
we plot the product $\tau_B D_{\alpha}$ for bonds in the surface and in
the interior. In agreement with the results from the bulk we find that
for shell~1 this product is independent of $T$ for the case of oxygen,
which shows that the breaking of the bond is the elementary diffusion step
for the oxygen atoms. This is not the case for the atoms in the surface,
since for silicon as well as oxygen the product depends on temperature
in that it decreases with decreasing $T$. Thus we conclude that the
breaking of a Si-O bond is not the elementary step for the diffusion
at the surface. One possibility for the explanation of this trend is,
i.e., that with decreasing temperature the number of dangling oxygens
is decreasing (see Fig.~\ref{fig8}) and hence the number of more mobile
atoms is decreasing. We also mention that Litton and Garofalini have
presented results from simulations in which they found that the enhanced
diffusivity of the particles near the surface is related to the fact
that these atoms can make larger jumps~\cite{litton97}. However, these
results were obtained at a relatively high temperature, i.e. in a regime
where the surface still has many arms sticking out into the vacuum (see
Fig.~\ref{fig1}a). Since these structures are not present at the lower
temperatures it should not be expected that this type of reasoning is
applicable also at low temperatures.

\begin{figure}[h]
\psfig{file=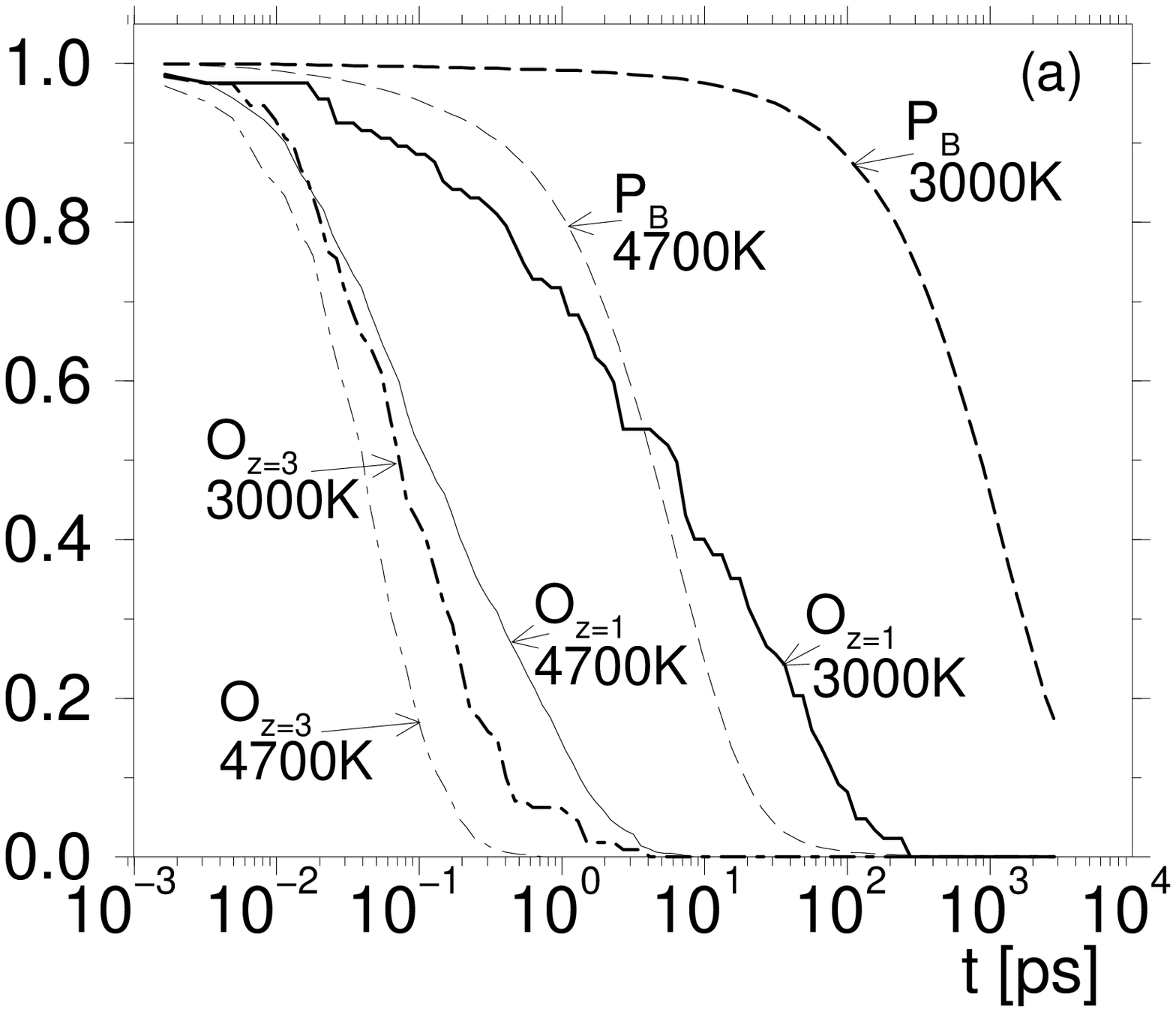,width=6.5cm,height=5cm}
\vspace*{5mm}
\psfig{file=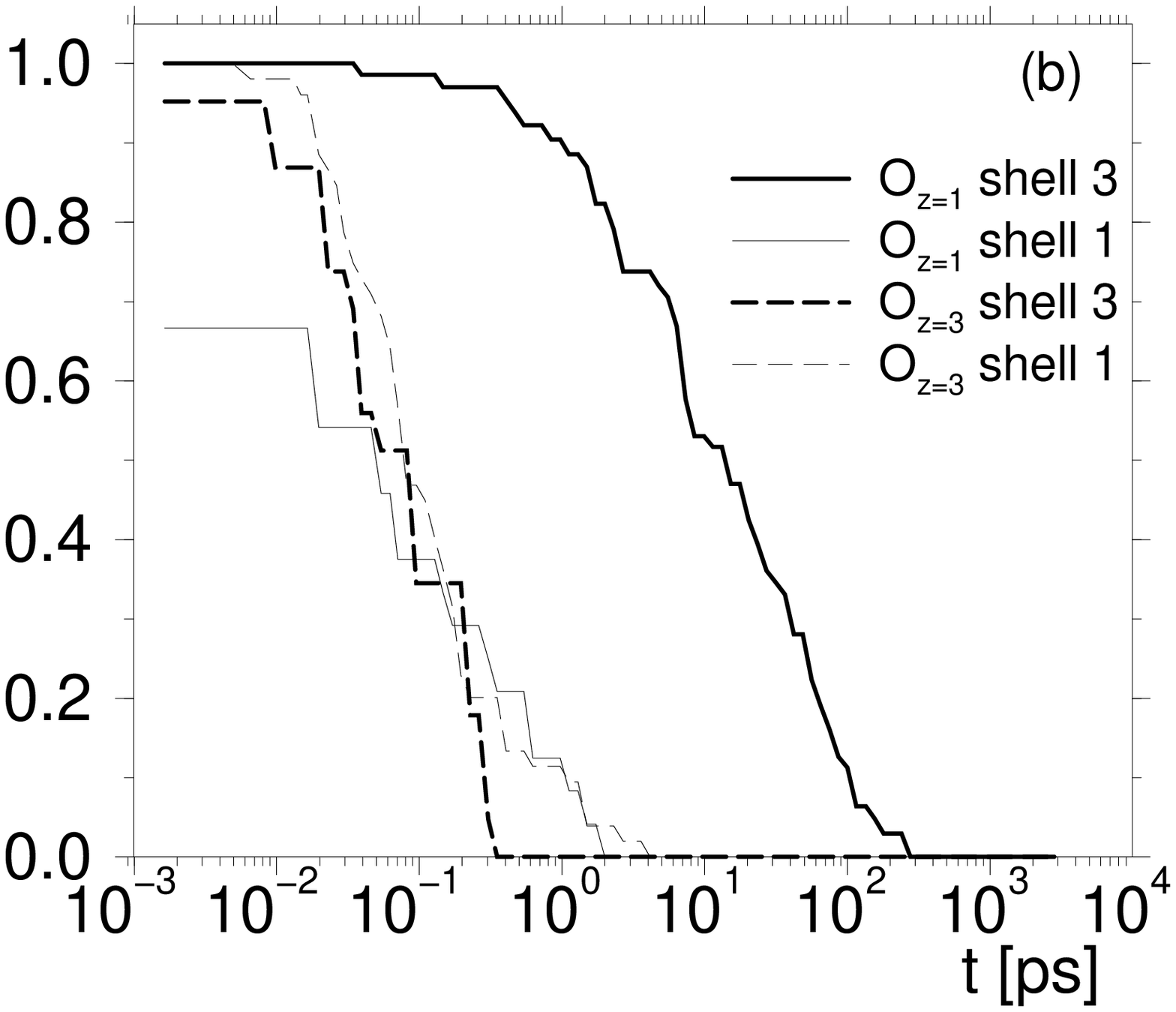,width=6.5cm,height=5cm}
\vspace*{5mm}
\caption{Probability that a bond (curves labeled $P_B$) and an oxygen defect
(curves labeled with $O_{z=1}$ and $O_{z=3}$) survives a time $t$. a) Whole system
for two different temperatures. b) Dependence on the shell for $T=3000$K.}
\label{fig15}
\end{figure}

It is also instructive to investigate the typical life-time of a
particular defect. For oxygen, the vast majority of defects are one and
three coordinated atoms (i.e. the coordination number is $z=1$ or $z=3$)
and thus we will concentrate in the following on these type of defects.
In Fig.~\ref{fig15}a we show for two temperatures the probability that
a defect that was present at time zero exists also at time $t$. From
this graph we recognize that the typical life-time for a defect of type
$z=3$, dashed-dotted curves, is much shorter, 0.1-1.0ps, than the one of
a bond, dashed curves. Furthermore we see that the temperature dependence
of this life-time is relatively weak, since a change of $T$ from 4700K
to 3000K gives rise to an increase of only a factor of 2-3, whereas the
curves for $P_B(t)$ are shifted by around a factor of 100. The situation
is very different for defects of type $z=1$, i.e. the dangling bonds
(solid curves).  Here we see that the curve shows a strong temperature
dependence, although it is not quite as strong as the one for $P_B(t)$.

The curves shown in Fig.~\ref{fig15}a are for the largest cluster,
taking into account {\it all} the atoms. In Fig.~\ref{fig15}b we now
study the dependence of these probabilities on the shell in which the
defect is for the case $T=3000$K. From this graph we recognize that the
life-time of a $z=3$ defect is very short in the bulk as well as on the
surface (dashed curves).  (Note that the statistics of these two curves
is very poor, since at $T=3000$K there are only very few defects of this
type.) For the dangling bonds, however, there is a big difference in the
survival probability (solid curves).  In the interior the defects are
annihilated very quickly, i.e. within 1ps, whereas on the surface they
survive for a long time. Thus we conclude that the large annihilation
time for the dangling bonds in Fig.~\ref{fig15}a is due to the fact that
on the surface these defects survive for a long time. The reason for this
large difference is probably that from an energetic point of view it is
{\it in the bulk} very unfavorable to have a defect and thus it is very
unstable. The situation is different on the surface, since a dangling
bond that sticks out from the surface does not cost too much energy and
thus can survive for a long time. Finally we mention that we found that
also the defect coordinated silicon atoms, mainly $z=3$ or $z=5$, stay in
this state only for a relatively short time, i.e. less than one ps. The
only exception are the three-fold coordinated Si-atoms at the surface
which can stay in that state for up to 2 ps~\cite{roder_diss}. But as
can be seen from Fig~\ref{fig15}b, even this time is much shorter than
the typical life-time of a dangling bond at the surface.

The final quantity we investigate is $g(\nu)$, the vibrational density
of state (DOS). For this we quenched a liquid sample from 3000K rapidly
to $T=300$K (100.000 steps = 163~ps) and, after annealing the sample
a bit, calculated $g(\nu)$ via the Fourier transform of the velocity
autocorrelation function~\cite{dove93}:

\begin{equation}
g(\nu) = \frac{1}{Nk_BT}\sum_j m_j
\int_{-\infty}^{\infty} dt \exp(i2\pi\nu t) \langle \vec{{\bf
v}}_j(t)
\vec{{\bf v}}_j(0) \rangle \quad .
\label{eq4}
\end{equation}

\begin{figure}[h] 
\psfig{file=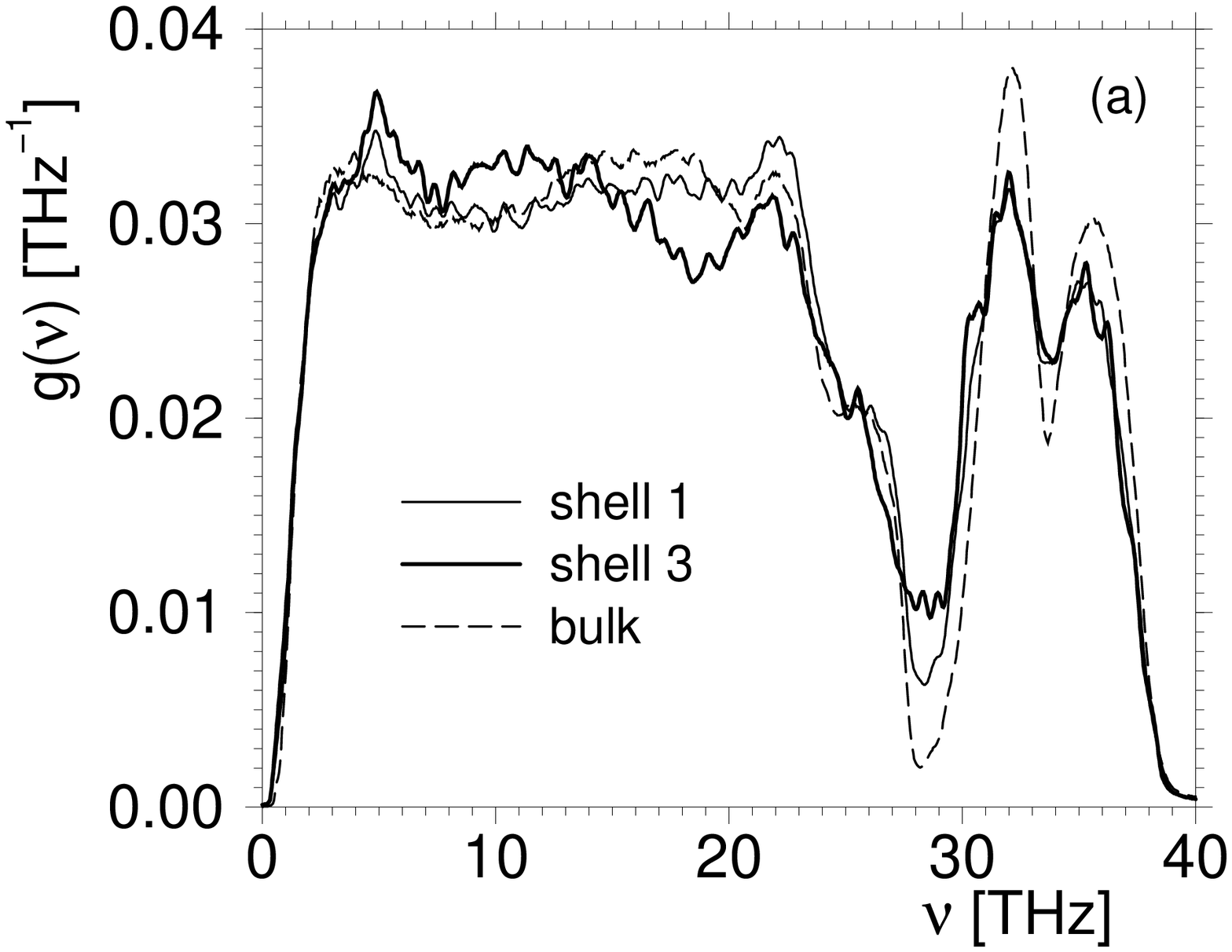,width=13cm,height=10cm}
\vspace*{5mm}
\psfig{file=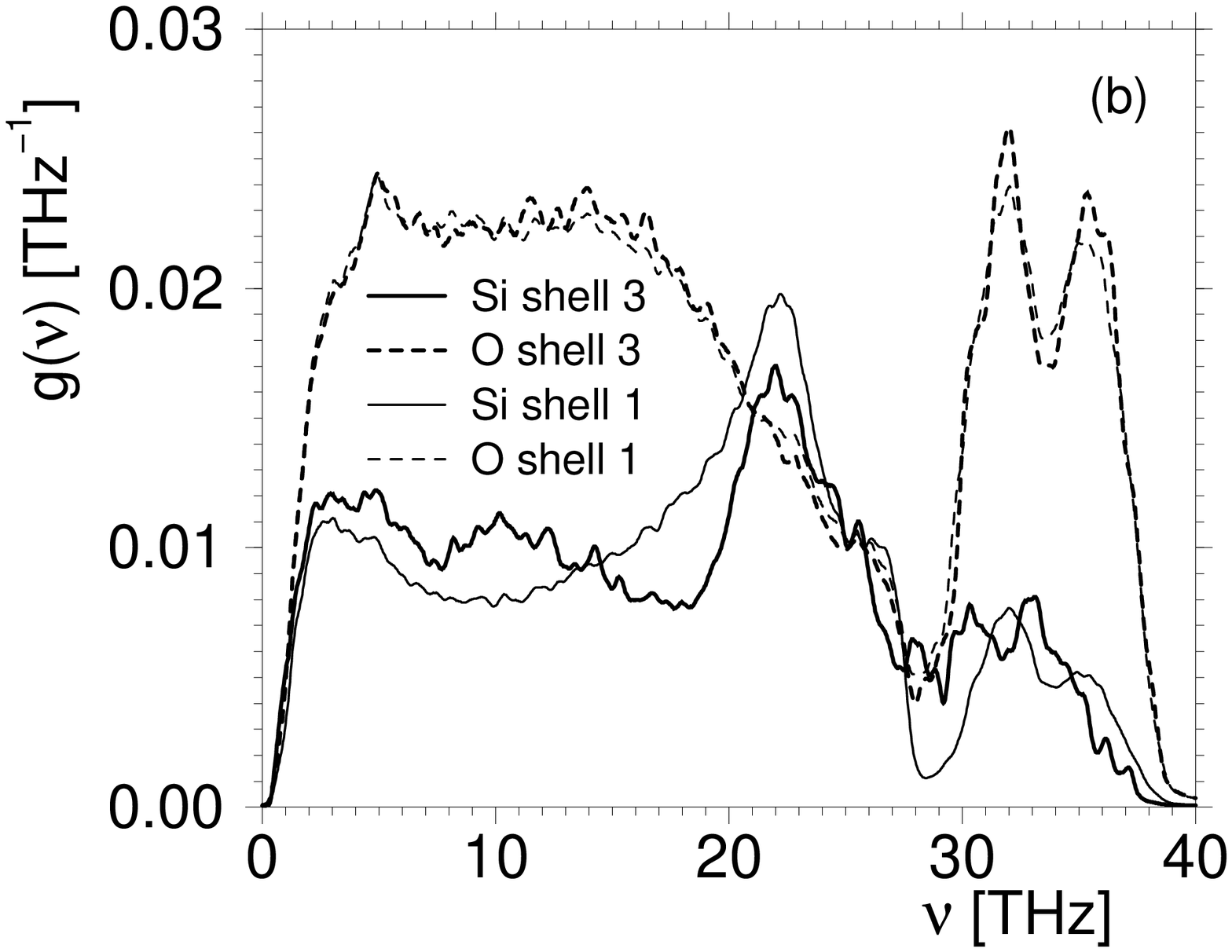,width=13cm,height=10cm}
\vspace*{5mm}
\caption{Frequency dependence of the density of state. a) The atoms in shell~1
and 2. Also included are the bulk data from Ref.~\protect\cite{horbach99c}.
b) Contribution to the DOS from silicon and oxygen atoms in shell~1 and shell~3.}
\label{fig16}
\end{figure}

The Fourier transform was calculated by making use of the Wiener-Khinchin
theorem~\cite{press86}. Note that this way to calculate the DOS allows
to decompose $g(\nu)$ into the individual contributions from the
different atoms, and hence into the atoms in the different shells. In
Fig.~\ref{fig16}a we show the so obtained DOS as obtained by considering
only the atoms in the interior, or in the surface layer (thin and bold
solid curves, respectively). We see that the two distributions are quite
similar and also quite close to the one of the bulk~\cite{horbach99c}. The
main difference is that in the curves for the cluster the double peak at
high frequencies is not separated as well as in the bulk and that also
the minimum around 28THz is not so pronounced. (We remind the reader
that the origin of these two peaks is an intra-tetrahedral motion of
the atoms~\cite{galeener79,pasquarello98}.) Since all three curves are
quite similar one might conclude that the presence of the surface does
not affect significantly the DOS. That this conclusion is incorrect is
shown in Fig.~\ref{fig16}b, where we show the contributions to the DOS
from the silicon and oxygen atoms in shell~1 and shell~3. From the graph
we recognize that the vibrational spectrum for the oxygen atoms is indeed
almost independent of the shell (dashed lines). In particular we see that
the location and the intensity of the two peaks at high frequency is
very similar in the two distributions. For the silicon atoms, however,
we find a significant difference between the two DOS. In the $g(\nu)$
for the surface the two peak at high frequencies are shifted from 32.0 and
35.2THz to 30.4 and 32.9THz, respectively. Since these peaks stem from the
motion of the atoms within one tetrahedron it is quite reasonable to find
that close to the surface this type of motion has a lower frequency, since
the atoms are less constrained. A further large change in the DOS is found
around 22.2THz, where on the surface the peak is less pronounced. Such
a decrease has also been found in the simulations by Beckers and de
Leeuw of porous silica~\cite{beckers00}, i.e. a system with a large
surface area. Since in that simulation the silica potential proposed by
Vashishta {\it et al.} was used~\cite{vashishta90}, we conclude that the
described decrease is not an artifact of the BKS potential but a real
dynamical feature of the silica surface. Experimental infrared studies
have shown that the presence of two-membered rings gives rise to bands
at 26.6 and 27.2THz~\cite{morrow76,bunker89a,grabbe95}. Although
this frequency is reproduced reasonably well by density functional
approaches~\cite{ceresoli00} which underestimates the two frequencies
by 5\%, the classical simulation have so far not been able to get a good
agreement with the experimental results. Although we find a small peak at
around 28THz, see Fig.~\ref{fig16}b, it is presently not clear whether it
does indeed correspond to the motion of the atoms in two-membered rings.

\section{Summary and Discussion}

The goal of this paper was to investigate the static and dynamical
properties of the surface of amorphous silica {\it in equilibrium}. For
this we did molecular dynamics computer simulations, using a potential
that in the past has been shown to be very reliable in the bulk. Two
types of geometries were investigated: Finite clusters with different
numbers of atoms (and hence surface area) and a thin film. Although
the results discussed in the present paper referred almost exclusively
to the clusters, we have found that the surface properties of the thin
film are, within the accuracy of our data, the same as the ones for the
clusters. Hence we conclude that the curvature of the surface in the
cluster geometry is basically irrelevant.

We have found that in the temperature range investigated the shape
of the clusters does not depend on $T$. From the size dependence of
the potential energy per particle we estimate that the surface energy
of the system is around 1 N/m, a result which is in rough agreement
with the experimental value for the surface tension. In agreement with
previous simulations~\cite{litton97} we find an enrichment of oxygen
close to the surface. The reason for this effect is that in this way the
number of energetically frustrated bonds is minimized. The study of the
partial radial distribution functions $g_{\alpha\beta}(r)$ as well as of
the distribution functions for the bond-bond angles shows that at the
surface there are structural units that are not existent in the bulk
(or rather they occur with an extremely small probability) and which
should be present also at low temperatures. By calculating $P(n)$,
the probability that a ring has length $n$, we show that short rings
are much more frequent on the surface than in the bulk. By determining
the structural properties of these short rings we demonstrate that
the mentioned anomalous features in the $g_{\alpha\beta}(r)$ and the
bond-bond angle distribution functions are closely related to these short
rings. The average density of the two-membered rings is 0.6 rings/nm$^2$,
in reasonable good agreement with experimental estimates. Also the
geometry of these two-membered rings is in good agreement with {\it ab
initio} simulations. Thus we come to the conclusion that the BKS-potential
used here is indeed able to give a realistic description not only of
vitreous bulk silica but also of the surface of amorphous silica. This
conclusion is at odds with the one of Ref.~\cite{ceresoli00} where it is
stated that this model gives rise to too many over and undercoordinated
surface atoms. Although the BKS-potential is certainly far from being
perfect the present work shows that if the surface is properly relaxed
the resulting structure is very reasonable.

We have also investigated the dynamics of the system. For times larger
than the ballistic regime we find that the MSD for the atoms at the
surface is larger than the one in the interior. Thus we conclude that at
the surface the cage of the atoms is larger than the one in the bulk. The
cluster-averaged diffusion constant depends significantly on the size
of the cluster and this dependency becomes stronger with decreasing
temperature. In order to study the diffusion mechanism in more detail we
have determined the probability that a oxygen defect which is present at
time zero is also present at time $t$. We find that an oxygen atom that
is three-fold coordinated stays in that state for less than a ps (whether
or not the defect is at the surface). In contrast to this the singly
coordinated oxygen atoms, i.e. the dangling bonds, stay in that state
for a very long time (100ps at 3000K) if the defect is at the surface,
whereas it disappears already after 1ps if it is in the bulk. Thus this
shows that on the surface the dangling bonds are very stable structures,
at least at these high temperatures.

Finally we have also determined the vibrational density of states
(DOS). We find that the total DOS for the whole cluster is quite similar
to the one of the bulk. However, if the total DOS is decomposed into the
parts corresponding to the silicon and oxygen atoms, we find that the
DOS for the silicon atoms close to the surface is quite different from
the one in the bulk. No such difference is found in the oxygen atoms.
Thus we conclude that the presence of the surface can be seen also in
the vibrational dynamics of the ions.

In summary we thus conclude that the BKS model seems to be capable to
give also a very reasonably description of the surface of amorphous
silica. Unfortunately not much {\it structural} information on this type of
surface is presently known from experiments and hence it is difficult
to judge the quality of the present results on a {\it quantitative}
level.  However, using new microscopic techniques like an STM or an
AFM it should be possible to obtain in the near future experimental
data~\cite{radlein95,raberg98,gupta00}. Furthermore it will also be valuable to
compare the present results with the ones obtained with a method that is
more accurate, such as the one proposed by Car and Parrinello~\cite{marx00}.
Although the time window that can be accessed by this method is more than
thousand times smaller than the classical simulations of the present
work, such a comparison will nevertheless give useful information on
the reliability of the BKS-potential in the presences of a surface.

Acknowledgement: We gratefully acknowledge the financial support by the
SCHOTT Glaswerke Fond and the DFG under SFB 262. Part of this work was
done at the NIC in J\"ulich.

\newpage

\begin{table}[hbt]
\begin{center} \begin{tabular}{|c|c|c|c|c|c|}  
 & $\theta_{\rm{OSiO}} [^{\circ}]$
 & $\theta_{\rm{SiOSi}}$ [$^{\circ}$]  
 & $d_{\rm{SiSi}}$ [\AA]
 & $d_{\rm{OO}}$ [\AA] 
 & $d_{\rm{SiO}}$ [\AA] \\ 
\hline
 present sim. &  81.5\, (14) & 98\, (17) & 2.5\, (0.36) & 2.28 \, (0.33) &1.65\,
(0.26) \\ \hline
sim. \cite{feuston89}     & 85   &  102     &         &  2.2   &  1.67 \\ \hline
quant. \cite{kudo84}      & 88.5 &  91.5    &   2.39  &  2.33  &  1.67 \\ \hline
quant. \cite{okeeffe85}   &      &  91.3    &   2.38  &  2.32  &  1.66 \\ \hline
quant. \cite{bunker89a}   & 88.5 &  91.5    &   2.39  &  2.33  &  1.67 \\ \hline
quant. \cite{hammann97}   & 89.7 &  90.3    &         &        &  1.68 \\ \hline
quant. \cite{lopez00}     & 89   &  91      &   2.40  &        &  1.69 \\ \hline
exp. \cite{weiss54}       & 100.4&  79.6    &         &        &  1.84 \\ 
\end{tabular}
\end{center}
\caption{Values for the various angles and distances in a two-membered ring
as determined from simulations and quantum mechanical calculations. In the
first line the numbers in brackets give the width of the distribution at
half maximum. The last line reports values from experiments on crystalline
silica with two-membered rings.}
\label{tab1}
\end{table}

\end{document}